# SOME PHYSICAL PROPERTIES OF A NEW JUPITER-FAMILY COMET P/2019 LD2 (ATLAS) FROM BROADBAND OBSERVATIONS

Serhii Borysenko[1], Gulchehra Kokhirova[2], Firuza Rakhmatullaeva[2]

[1]Main Astronomical Observatory of NAS of Ukraine, Akademika Zabolotnoho 27, Kyiv 03143, Ukraine

[2]Institute of Astrophysics of NAS of Tajikistan, Bukhoro 22, Dushanbe 734042, Tajikistan

e-mail: borisenk@mao.kiev.ua

## ABSTRACT

New periodic comet P/2019 LD2 (ATLAS) located on unstable quasi-Trojan orbit still an interesting object to study in the last years. We present the results of broadband observations of comet P/2019 LD2 (ATLAS) performed at the Sanglokh observatory of the Institute of Astrophysics, National Academy of Sciences of Tajikistan for 5 nights in August 2020. The dependence of the *Afρ* parameters of the comet on the aperture radius is measured. We obtained the coma corrected upper limit of the cometary radius $R_n \leq 6.1 \pm 0.1$ km and calculated the absolute magnitude $H_0 = 11.41 \pm 0.03$. Finson – Probstein diagram method used to explain a dust tail appearance. A subgroup of comets with orbital parameters close to P/2019 LD2 (ATLAS) is analyzed.

INTRODUCTION

Comet P/2019 LD2 was discovered in images by the Asteroid Terrestrial-impact Last Alert System (ATLAS) at the Mauna Loa Observatory taken on 10 June 2019 (CBET 4780, MPEC 2020-K134). Follow-up observations by the Las Cumbres Observatory on 11 and 13 June 2019 confirmed the cometary appearance of P/2019 LD2, which now had a more apparent coma and tail. Later observations by the ATLAS-MLO in April 2020 showed that P/2019 LD2 still retained its cometary appearance, suggesting that it has been continuously active for almost a year.

P/2019 LD2 is a Jupiter-family comet with a Tisserand parameter of 2.94, typical for other Jupiter-family comets. The comet's nominal orbit suggests that it is not in a stable 1:1 resonance with Jupiter as it has made a close approach to the planet on 17 February 2017, at a distance of 0.092 AU (13.8 million km), and will make a similarly close approach in 2028. Unlike the Jupiter Trojans, P/2019 LD2 is 21 degrees ahead of Jupiter, and will continue drifting 30 degrees ahead before returning back to Jupiter and making close approaches (Steckloff et al. 2020; Kareta et al. 2020; Hsieh et al. 2021).

We selected some subgroup of comets with orbital parameters close to P/2019 LD2 comet (comets with semimajor axis between 4.7 – 5.3 AU and eccentricity less than 0.2) (Tab. 1). All of these objects are faint comets with observed tiny comae and tails close to the oppositional location and at perihelion (http://www.aerith.net). Some of them could be discovered at an advanced activity state (404P, P/2012 T2, P/2020 O3). The absolute magnitude of comets in this subgroup by MPC (Minor Planet

Center) data lay in the range of about 8 – 12 and Tisserand parameter distribution is 2.75 – 2.98. By these values, we can say that P/2019 LD2 is one of the typical comets in the list.

[Tab. 1]

A "quasi-body" is an object in a specific type of co-orbital configuration with a planet where the object stays close to that planet over many orbital periods. In the case of quasi – Trojan comets such objects have not stable dynamics and their location can change. One of the distinctive physical features of such comets may be the distribution of normalized reflectivity gradient S' for their nuclei which lay in the range 10 – 20 whereas for Trojans it in the range 0 – 10 (Jewitt et al. 2004).

Curves of brightness for comets are strongly dependent on the locations on the diagram (Fig. 1). Bodies located bottom of the diagram usually demonstrates only dust activity. On other hand, this is probably purely an observational bias. Those objects with low orbital eccentricity stay at large heliocentric distances, making detection of faint gas emission difficult. Objects with higher eccentricity, and therefore smaller perihelion distances, can be observed at closer distances to Sun and Earth, making detection of the increased gas emission rates much easier. The typical behaviour for the curve of brightness as for other objects in Table 1 is expected for P/2019 LD2 in the nearest future.

[Fig. 1]

Comets with temporary positions near Jupiter's orbit are interesting for study (Fig.1). The nominal orbits of these objects suggest that they are not in a stable 1:1 resonance with Jupiter. Numerical integrations show that these objects were previously Centaur before reaching their current Jupiter Trojan-like orbit and can return to Centaur orbit before eventually becoming a Jupiter-family comet in future. Active Centaurs, defined as solar system bodies whose orbits have both perihelia and semimajor axes between the orbits of Jupiter (at 5.2 AU) and Neptune (30 AU) and which are not in 1:1 resonance with the giant planets (Jewitt 2009a, 2009b; Sarid et al. 2019). The orbit of Comet P/2019 LD2 will remain approximately the orbit until 2028 (Fig.1). Then, is expected, the comet will return to a Centaur-like orbit in 2028 and remain there until 2063, when a very close encounter with Jupiter at ~0.03 AU in 2063 will lower both its semimajor axis and perihelion distance to well below $a_J$, at which point, the object will be considered a JFC (Hsieh et al. 2021). The case of P/2019 LD2 highlights the need for mechanisms to quickly and reliably dynamically classify and studying the physical properties of such bodies discovered in current and upcoming wide field surveys (Bolin et al. 2021; Steckloff et al. 2020).

Steckloff et al. (2020) found that LD2 is most likely a pristine comet. Although it has likely lost some supervolatile ices such as carbon dioxide ice in the outer solar system beyond Jupiter, it is unlikely to have ever been in the inner solar system, which is warm enough for water ice to sublime. The comet presents a unique opportunity to observe how pristine JFCs behave as their water ice begins to sublime for the first time and drive comet activity. Moreover, this transition is likely to finish in only 40 years from now. This means that people alive today will be able to follow this object all the way through its final transition into the JFC population.

OBSERVATIONS

Observations of P/2019 LD2 were made at the Sanglokh observatory, Tajikistan (MPC code – 193), with the 1-m Zeiss-1000 telescope and FLI Proline PL16803 4096×4096 9 μm CCD sensor with readout noise –10 e⁻ (estimated gain – about 0.497 e⁻/ADU). The camera is installed at the telescope's Cassegrain focus (F = 13 300 mm) and equipped by Johnson – Cousins broadband filters *R* and *V*. Binning 4x4 was used, resulting in an effective pixel scale of 0.579"/pixel and FOV about 10′ × 10′.

Observations were made in photometric sky conditions without Moon or with low elevations of Moon (< 7° in phase 0.8 – 0.9). Average seeing (FWHM of faintest stars) was 1.2″ – 2″.

Image processing and stacking the frames was done with the *Astroart* software (http://www.msb-astroart.com/). We used ATV IDL (Barth et al. 2001) routines for aperture photometry of comets and reference stars.

The APASS-DR9 star catalog was used as a photometric reference (Henden et al. 2016). This catalog includes magnitudes of stars from about $7^{th}$ magnitude to about $17^{th}$ magnitude in five filters: Johnson *B* and *V*, plus Sloan *g′*, *r′*, *i′*. It has mean uncertainties of about 0.07 mag for *B*, about 0.05 mag for *V* and less than 0.03 mag for *r'*. We used the transformation formula from *r'* to $R_C$ magnitudes derived by Munari et al. (2014).

A log of observations is listed in Table 2.

[Tab. 2]

DISCUSSION

*The dust production parameter*

The *Afρ* parameter was calculated from (A'Hearn et al. 1984):

$$Af\rho = 4r^2\Delta^2 \cdot 10^{0.4(m_{Sun}-m_c)}\rho^{-1} \quad , \quad (1)$$

here *r* (AU) is the heliocentric distance; Δ (cm) is the geocentric distance; $m_{Sun}$, $m_c$ are the apparent *R* magnitudes of the Sun and of the comet, respectively; ρ [cm] is the radius of the photometric aperture projected onto the sky.

For observations obtained with the 1-m Zeiss-1000 telescope, the *Afρ* parameter was calculated using images obtained in the *R* filter and with the aperture projection radius ρ = 10588 km (4.05″). The aperture radius corresponds to the maximum of the signal-to-noise ratio determined by the curves-of-growth analysis of each night of data using *Astroart* software. The values of the dust productivity parameter, along with the magnitude of the absolute brightness given below, also confirm the increased activity of the comet, in addition, during the observation period, the heliocentric distance gradually increased, in this regard, a tendency to decrease in the parameter is observed. Figure 2 shows the distribution of the parameter *Afρ* depending on the radius of the measurement aperture, found from our observations. Obtained *Afρ* values are large than for typical quasi-Trojan comets 158P/Kowal – LINEAR and 244P/Scotti presented in the database of observations of comets and asteroids by Spanish amateur and professional astronomers (http://www.observadores-cometas.com/). Observed *Afρ* data for both of the objects are less than 30 cm. For comets from others, neighbor orbital groups, such as quasi-Hilda comets, *Afρ* values are usually less than 100 cm (except for the brightest objects) (Borysenko et al., 2019; 2020a). Kareta et al. (2021) also find thus comet P/2019 LD2 appears more active than a JFC at a comparable distance.

Bolin et al. (2021) showed *Afρ* estimations for the comet P/2019 LD2 between 147.13 – 192.88 cm in August 2020 by observations in the *r* filter with aperture radius $\rho$ = 10 000 km. These values are normalized to 0° phase angle and slightly less than our values obtained by *R* magnitudes (Table 2). The differences can be explained mostly by different filters used and some better seeing.

Future observations are useful to study the comet as a transient object that has likely never entered the inner Solar System for any significant period of time (Kareta et al. 2021).

[Fig. 2]

*Absolute magnitude and size of cometary nucleus*

We used apparent magnitudes $m_R$ to obtain the absolute magnitude of the comet using the equation:

$$H_R = m_R - 5\log(r\Delta) - \beta\alpha \quad , \qquad (2)$$

in which $r$ and $\Delta$ are the heliocentric and geocentric distances, respectively; $\alpha$ is a phase angle, and $\beta$ [mag/deg] is the linear phase coefficient (assumed to be 0.04, Lamy et al. 2004) equal to the ratio of flux densities scattered at angle $\alpha$ to $\alpha = 0°$.

The obtained average value of absolute magnitude is $H_R = 11.41 \pm 0.03$ for the dust coma within 4.05″ aperture radius. Bolin et al. (2021) find absolute magnitude for the cometary nucleus in *V* filter as $H_V = 15.53 \pm 0.05$ by results of photometric observations in April 2020.

A rough estimation of upper limit of radius for cometary nucleus was calculated as:

$$R_n^2 = 2.238 \cdot 10^{22} r^2 \Delta^2 10^{0.4(m_\odot - m_{nucl} + \alpha\beta)} A^{-1} \quad , \qquad (3)$$

here $r$ [AU] is the heliocentric distance; $\Delta$ [AU] is the geocentric distance; $m_\odot$, $m_{nucl}$ are the apparent *R* magnitude of the Sun and a nuclear magnitude of a comet respectively; $\alpha$ [deg] is the phase angle and $\beta = 0.04$ [mag/deg] – phase coefficient (Russel, 1916). The geometric albedo value A = 0.12 suggested by Fernández et al. (2009) was used. A rough estimation of radius on August 8, 2020 gives value for nuclear radius $R_n \leq 7.4 \pm 0.1$ km.

Kareta et al. (2021) and Bolin et al. (2021) both find evidence for relatively stable activity of cometary dust ejection speeds for August 2020, making any changes in brightness rather gradual. We analyzed the profile of comet for one of the best seeing day during our observations (August 8, 2020) to obtain a more appropriate, coma corrected value for nuclear magnitude and the size of the nucleus. We applied the coma compensation method described in Hicks et al. (2007) and references therein. In this method, a stellar point spread function (PSF) obtained from background stars is used to fit the innermost coma. The coma was assumed to dominate the skirt between the curves (Figure 3). The profile of the coma taken along the normal direction to the cometary tail to avoid the influence of the tail on the modeled coma profile. Star PSF was created and averaged by reference stars with a magnitudes close to cometary nucleus estimation, has FWHM 5.65 pix (about 3.3 arcsec) and averaged for X and Y profiles. The ratio of the areas under stellar PSF and total coma is proportional to the nuclear contribution and the total flux ratio.

As shown in Fig. 3 the image of comet P/2019 LD2 was low dominated by coma on August 8, and we obtained coma corrected value of $m_{nucl}$ = 18.14 mag. Thus, the radius of the nucleus for comet P/2019 LD2 (ATLAS) calculated by Formula 2 is $R_n \leq 6.1 \pm 0.1$ km.

[Fig. 3]

Bauer et al. (2013) suggest the distribution of Centaur albedos between ~5 and ~11 percents groups. Therefore, the nucleus of comet P/2019 LD2, as Active Centaur, is more likely to be darker and our estimations for radius can value between 9.4 ± 0.2 km and 6.4 ± 0.1 km. Kareta et al. (2021) obtained the values for cometary radius 5.1 ± 0.1 km assuming a 5% geometric albedo, or 3.4 ± 0.1 km assuming an 11.2% albedo by *g'* photometry with DECam (Dark Energy Camera) installed on the Cerro Tololo Inter-American Observatory's Blanco 4 m telescope. Precovery observations from the DES's (Dark Energy Survey, Abbott et al. 2018) DECam instrument (Flaugher et al. 2015) from 2017 March 6 find no detectable object in the vicinity of where P/2019 LD2 should be down to $r' \sim 23.8$, which suggests the radius of comet to be ~1.2 km or smaller given a 5% visible albedo. Bolin et al. 2021 indicate that the nucleus of P/2019 LD2 has a radius between 0.2 – 1.8 km assuming a 0.08 albedo.

We obtained color index $V - R = 0.51 \pm 0.09$ for comet P/2019 LD2 on August, 6 (Table 2). According to Jewitt (2009, 2015), color index $V - R$ for Active Centaurs 0.50 - 0.51, for Active JFC – 0.46 - 0.47 (Solontoi et al. 2012), and for Jupiter Trojan – 0.45 (Szabó et al. 2007) (obtained by $g - r$ data using the photometric transformations from Ivesić et al. 2007). Most Active Centaurs are slightly redder than the Sun at optical wavelengths, but physical interpretation of the coma colors is difficult due to potential gas contamination (Jewitt 2009).

*Finson – Probstein diagram*

To explain the direction and length of the tail for cometary objects we can use Finson – Probstein diagrams (Finson & Probstein 1968), presented in celestial coordinates and compared with the real image of the object (Fig. 4). Finson & Probstein (1968) proposed a model which describes the full tail geometry with a grid of synchrones and syndynes. This model is simple because it considers only particles released in the orbital plane of the comet, and with zero initial velocity, but it provides a very good approximation of the shape of the tail, and has been used successfully to study cometary tails (Kramer et al. 2017; Chu et al. 2020; Borysenko et al. 2020b).

In the tail, dust and gas are decoupled and the only significant forces affecting the grain trajectories are the solar gravity and radiation pressure $P_{radiation}$. Both forces depend on the square of the heliocentric distance but work in opposite directions. Their sum can be seen as a reduced solar gravity, and the equation of motion is simply

$$m \times a = (1 - \beta) \times g_\odot, \qquad (4)$$

where $m$ is the mass of the dust particle, $a$ is the acceleration of the dust particle, $\beta$ is the ratio $P_{radiation}/g_\odot$, and is inversely proportional to the size of the grains for particles larger than 1 $\mu m$ (Vincent 2014).

Synchrones and syndynes represent respectively the locations of particles released at the same time, or with the same $\beta$ (Equation (4)). A range of $\beta$ values (0.005, 0.010, 0.015, 0.020, 0.025, and 0.03) were used to generate the syndynes shown with time of emissions 50, 100, 150, and 200 days for synchrones with integration step 1 day. The curves span the width of the dust tail as seen in Fig.4. Taking into account that $\beta = 0.57 Q_{pr}/\rho a$, where $\rho$ is the density of the dust grain, expressed in grams per cubic centimeter, $a$ is the radius of the dust grain, in micrometers, $Q_{pr}$ is the efficiency of the radiation pressure, which depends on the size, shape and optical characteristics of a dust grain (for cometary dust, the radiation pressure efficiency is usually about 1), then, taking the density of cometary dust to be about 0.1 g/cm$^3$ (Greenberg & Li 1999), we can obtain information about sizes of the particles in the coma and tail (Kelley et al. 2013). It can be said that large particles > 100 μm in size dominate the inner regions of the coma for comet P/2019 LD2 (Fig. 4). It is in agreement with Kareta et al. (2021), but our data might not be sufficiently deep to detect smaller grains from this kind of modeling, but that doesn't mean that they don't exist.

[Fig 4.]

*Morphological study and mathematical filtering*

For best visualization of dust tail and some dust structures in cometary coma (jets, shells, fans) using of some mathematical filtering can be useful. Some methods enhance the radial tail, others reveal the azimuthal arcs, and some remove the azimuthal variations, while in others it is retained. Although dramatically different results are obtained with different techniques, it is clear that no single enhancement is ideal for every situation (Schleicher & Farnham 2004). This emphasizes why multiple techniques should be used to investigate the variety of potential coma features that may be present (Fig. 5).

[Fig. 5]

Radial weighted model (RWM) filter help to subtract coma from the image and specify the location of the cometary nucleus. This method based on Bonev & Jockers (2002). The algorithm extracts the pixel values from the comet coma, subtracts the background value and multiplies for the cometocentric distance to create a new image with the computed values. This procedure removes the mean $1/\rho$ gradient in the images and enhances the radial structures.

RWM filter emphasizes some tail bend in the inner part of the cometary atmosphere and some low asymmetric location of a cometary nucleus for P/2019 LD2. On the other hand, Larson-Sekanina (LS) filter accents of the near nuclear region and show the comma-like shape of the innermost comae. LS filter is usually used to separate and visualize dust structures in gas contaminated atmospheres and it is not such effective for purely dusty atmospheres.

Using of divide by azimuthal median filter (Samarasinha & Larson 2014) can help to find different asymmetric structures in the inner coma. Using the filter is useful for the detection of outbursts (like jets) in the nearnuclear regions. In our case, the filter shows an excess of coma material in the tail direction of the comet. Thus, the results of the morphological study show relative stability in the inner coma of comet P/2019 LD2 (ATLAS). However, Bolin et al. (2021) report about a ~1″ jet-like structure is seen with a position angle of ~210° from images of P/2019 LD2 taken with HST/WFC3 on 2020 April 1 UTC.

CONCLUSIONS

Photometric observations of comet P/2019 LD2 (ATLAS) obtained in August 2020 are presented. We obtained the absolute magnitude of the comet in the $R$ band is equal to 11.41 ± 0.03 mag. The calculated $Af\rho(\rho)$ relation is plotted in Fig. 2 for the projected distances from about 1500 km to 26 000 km from the nucleus.

A simple PSF-fitting model was used to estimate coma contamination. We obtained the coma corrected nuclear magnitude of the comet and upper limit of the cometary radius $R_n \leq 6.1 \pm 0.1$ km.

Finson – Probstein diagram analysis implies that the >100 μm-scale dust particles dominate the inner regions of the coma for comet P/2019 LD2.

Photometric data indicate that during the monitoring period, the comet was in a state of slightly increased cometary activity (relative to neighboring comets, Fig.1), associated mainly with the recent passage of perihelion. These data confirm the conclusion of Bolin et al. (2021) and Kareta et al. (2021), which showed that the region of the location of transit objects (from the Centaurs to comets of the Jupiter family), where P/2019 LD2 (ATLAS) belongs, coincides with the heliocentric range of distances, where the activity of the observed cometary bodies increases significantly. The comet's actual orbit indicates that the comet is in the transition from the Centaurs to the comets of the Jupiter family. The slightly increased $Af\rho$ value compared to other comets of the Jupiter family (Borysenko et al., 2019: 2020a) may indicate a previous long stay of the comet in the outer regions of the Solar System.

ACKNOWLEDGMENTS

Figure 1. Subgroups of short-periodic comets. Short-periodic comets signed by filled circles; comet P/2019 LD2 (ATLAS) signed by Astra symbol; Quasi-Trojan comets signed by filled triangles. The dashed line shows the location of the Jupiter orbit (1:1 resonance).

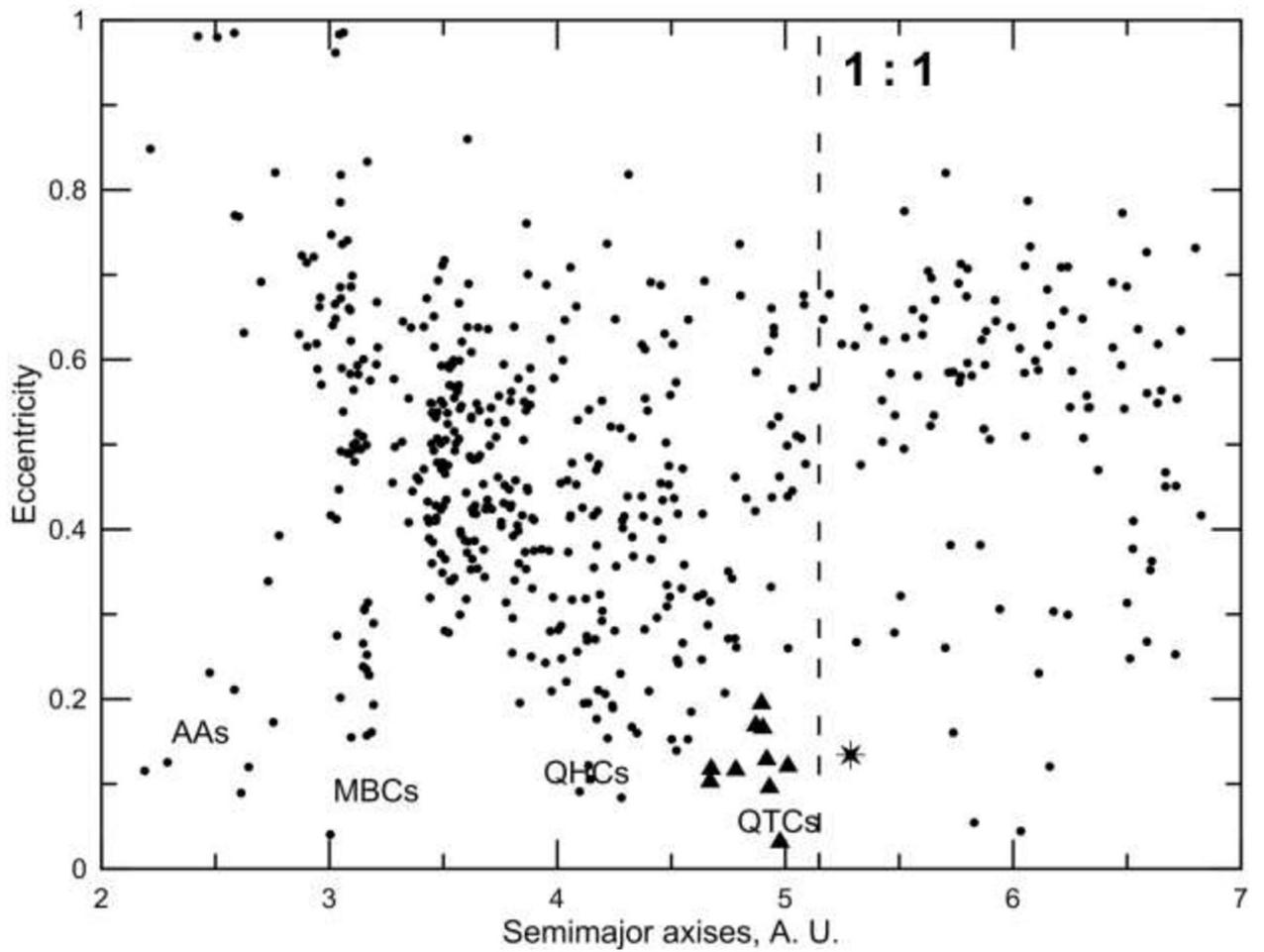

Figure 2. Distribution of the *Afρ* parameter depending on the radius of the photometric aperture.

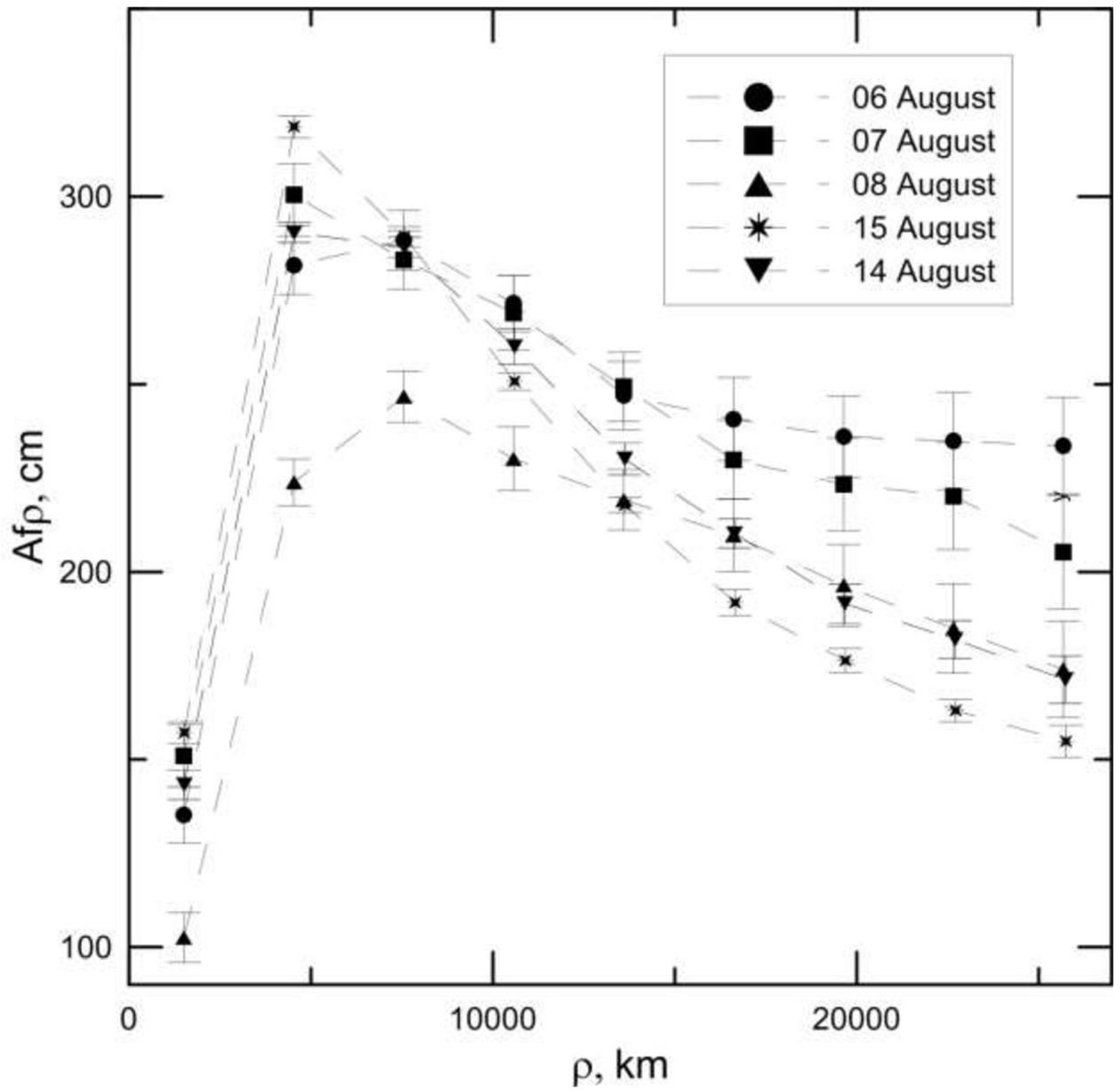

Figure 3. PSF fitting model to determine coma contamination. Stellar PSF averaged for X and Y profiles. In some areas, the skirt of the coma (circle symbols) stands clearly above the background sky (triangle symbols). This difference was used to model coma contribution.

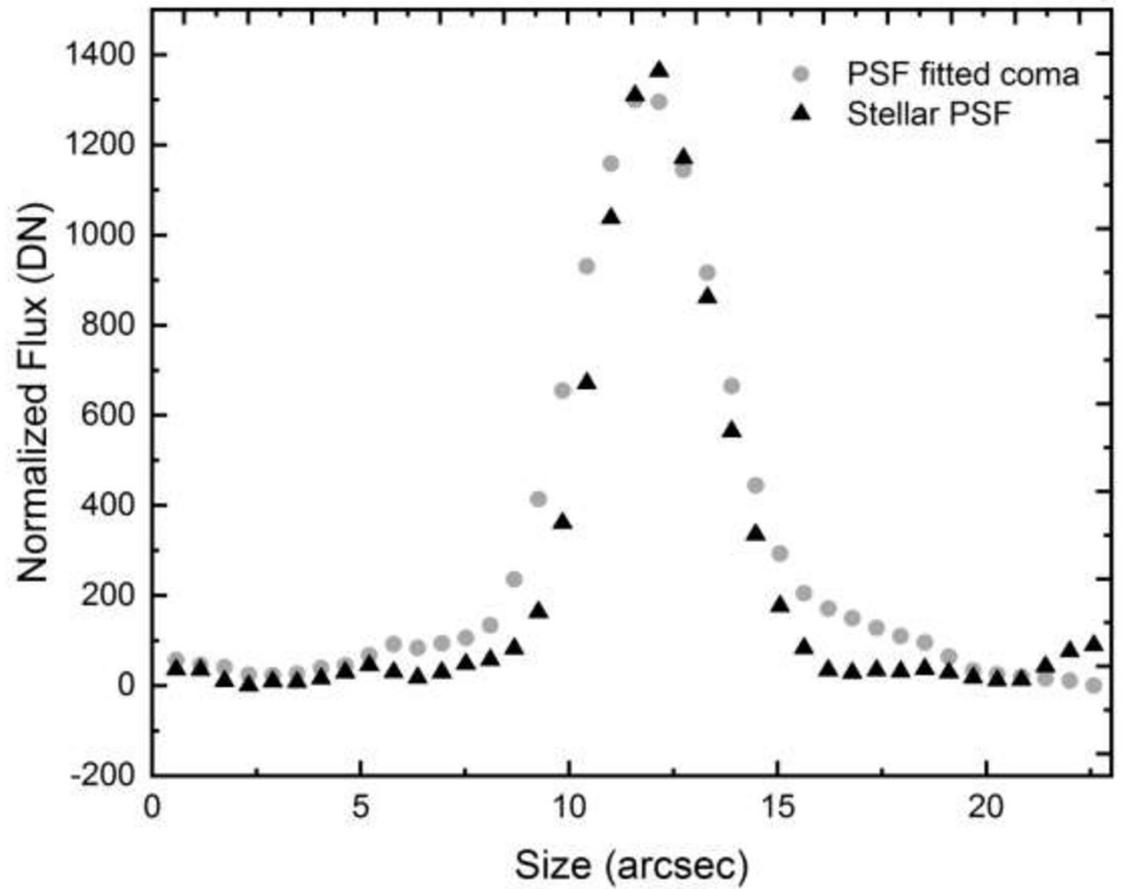

Figure 4. Finson – Probstein diagram of comet P/2019 LD2 (ATLAS) superimposed on a grid in real scale for original R-band image of comet P/2019 LD2 (ATLAS) obtained in 2020 (August 15, total expos. 53 x 120 s) with 1-m Zeiss-1000 telescope of the Sanglokh observatory, Tajikistan (Borysenko, Kokhirova, Rakhmatullaeva), showing the distribution of syndynes and synchrones throughout the visible part of the coma (Vincent 2014).

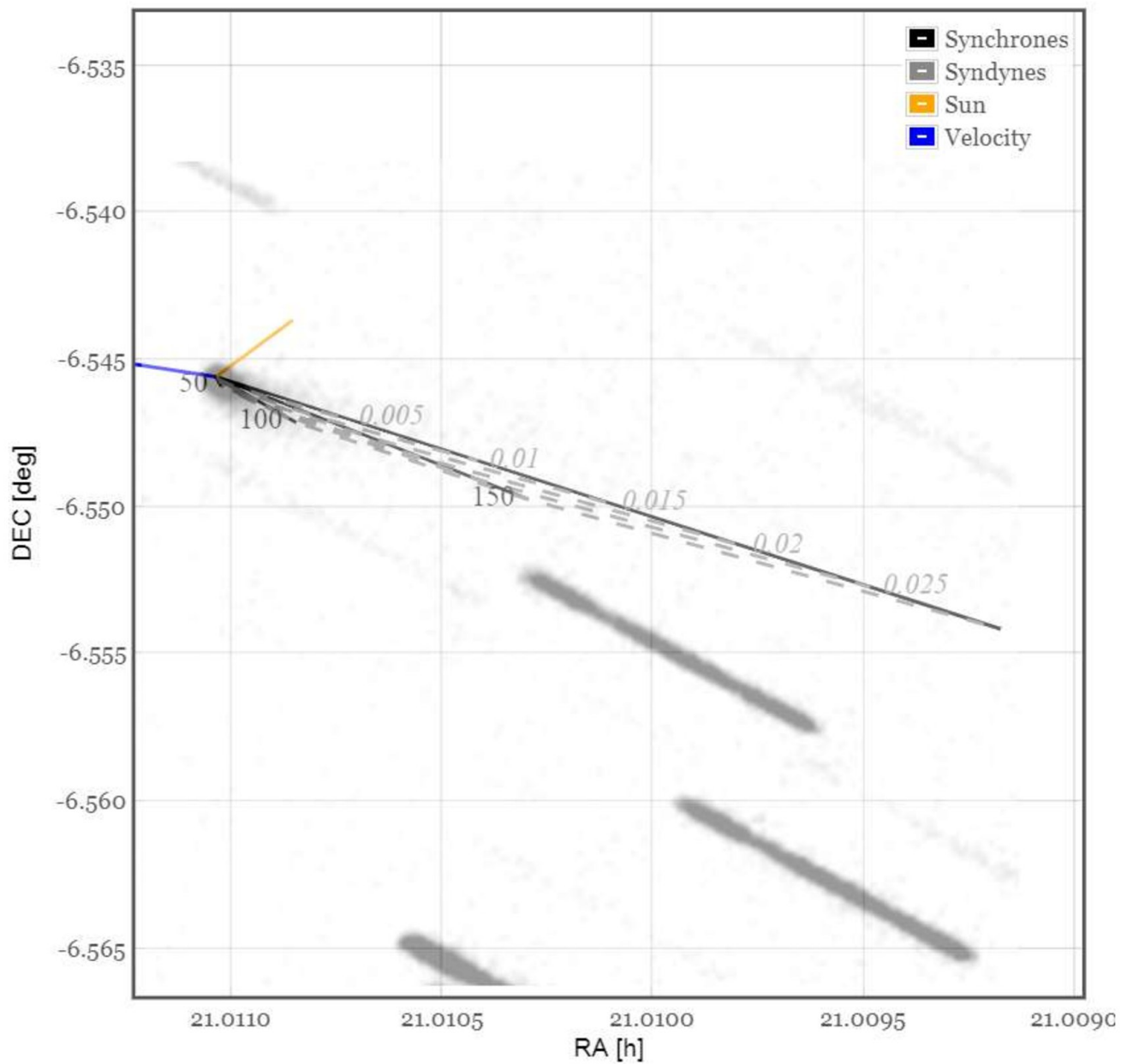

Figure 5. a) Original R-band image of comet P/2019 LD2 (ATLAS) obtained in 2020 (August 15, total expos. 53 x 120 s) with 1-m Zeiss-1000 telescope of the Sanglokh observatory, Tajikistan. The cometary tail was about 50″ at this time in an anti-motion direction; b) radial weighted model (RWM) filtered image; c) Larson–Sekanina ($\Delta R = 8$; $\alpha = 30°$) filtered image; d) dived by azimuthal median filtered image. Images obtained by using *Astroart* software.

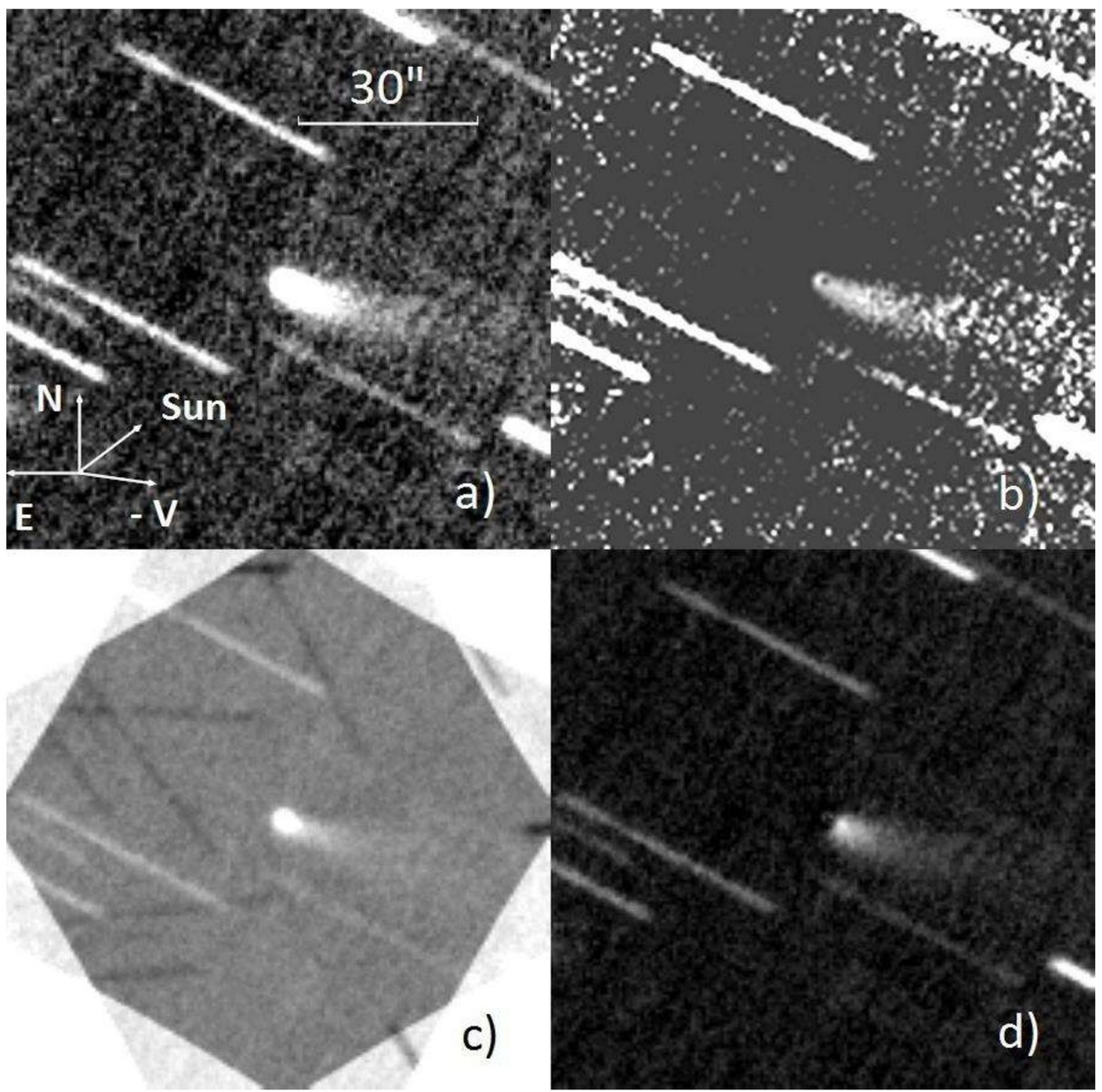

Table 1. Quasi-Trojan subgroup of JFC.

| Comet | $q^a$, AU | $e^b$ | $a^c$, AU | $i^d$ | $P^e$, (year) | $T_J^f$ | $H_0^g$ | Ref. |
|---|---|---|---|---|---|---|---|---|
| 158P/Kowal-LINEAR | 4.800 | 0.035 | 4.976 | 8.017 | 11.1 | 2.981 | 9.0 | MPEC 2016-K18 |
| 186P/Garrad | 4.389 | 0.124 | 5.012 | 28.513 | 11.2 | 2.750 | 7.5 | MPEC 2020-P19 |
| 244P/Scotti | 3.924 | 0.198 | 4.893 | 2.259 | 10.8 | 2.963 | 9.0 | MPC 88331 |
| 281P/MOSS | 4.034 | 0.172 | 4.872 | 4.720 | 10.7 | 2.968 | 11.0 | MPC 83145 |
| 391P/Kowalski | 4.111 | 0.121 | 4.675 | 21.276 | 10.1 | 2.867 | 8.0 | MPEC 2020-A105 |
| 393P/Spacewatch-Hill | 4.209 | 0.120 | 4.783 | 16.796 | 10.5 | 2.910 | 11.0 | MPEC 2020-P19 |
| 404P/Bressi | 4.268 | 0.132 | 4.919 | 9.973 | 10.9 | 2.956 | 10.0 | MPEC 2020-P19 |
| P/2004 FY140 (LINEAR) | 4.074 | 0.169 | 4.903 | 2.135 | 11.0 | 2.973 | 12.5 | MPC 54823 |
| P/2017 U3(Panstarrs) | 4.443 | 0.099 | 4.932 | 15.913 | 11.0 | 2.918 | 11.0 | MPEC 2020-P19 |
| P/2019 LD2 (ATLAS) | 4.578 | 0.135 | 5.294 | 11.552 | 12.2 | 2.941 | 8.5 | MPEC 2020-P19 |
| P/2020 O3 (PANSTARRS) | 4.175 | 0.106 | 4.671 | 8.445 | 10.1 | 2.978 | 12.0 | MPEC 2020-P19 |

[a]Perihelion distance.
[b]Eccentricity.
[c]Semimajor axis.
[d]Inclination of orbit.
[e]Orbital period.
[f]Tisserand parameter
[g]Absolute magnitude by MPC data

Table 2. Log of observations and results of measurements.

| Date, UT | $N_i \times Exp^a$ | $r^b$, AU | $\Delta^c$, AU | $\alpha^d$, (deg) | $R^e$, mag | $V^f$, mag | $V - R$ | $Af\rho^g$, (cm) | $\rho^h('')$ |
|---|---|---|---|---|---|---|---|---|---|
| August 06 17:03:33 | 28x120s | 4.591 | 3.590 | 2.3 | 17.53±0.03 | 18.04±0.09 | 0.51±0.09 | 272±7.5 | 4.05 |
| August 07 17:31:26 | 41x120s | 4.591 | 3.590 | 2.3 | 17.54±0.04 | – | – | 269±9.9 | 4.05 |
| August 08, 16:59:45 | 23x120s | 4.592 | 3.590 | 2.2 | 17.71±0.04 | – | – | 230±8.5 | 4.05 |
| August 14 21:43:15 | 24x120s | 4.593 | 3.597 | 2.6 | 17.58±0.02 | – | – | 260±4.8 | 4.05 |
| August 15 20:36:11 | 53x120s | 4.593 | 3.599 | 2.7 | 17.62±0.01 | – | – | 251±2.3 | 4.05 |

[a] Number of stacking images and exposure of the each one.
[b] Heliocentric distance.
[c] Geocentric distance.
[d] Phase angle.
[e] Apparent total red magnitude of the comet.
[f] Apparent total visual magnitude of the comet.
[g] $Af\rho$ parameter of the comet.
[h] Radius of photometric aperture.